# EEG fingerprinting: subject specific signature based on the aperiodic component of power spectrum

Matteo Demuru[1] and Matteo Fraschini[2]

[1] Stichting Epilepsie Instellingen Nederland (SEIN), Heemstede, The Netherlands
[2] Department of Electrical and Electronic Engineering, University of Cagliari, Cagliari, Italy

Corresponding author: Matteo Fraschini (e-mail: fraschin@unica.it).

## Abstract

During the last few years, there has been growing interest in the effects induced by individual variability on activation patterns and brain connectivity. The practical implications of individual variability is of basic relevance for both group level and subject level studies. The Electroencephalogram (EEG), still represents one of the most used recording techniques to investigate a wide range of brain related features. In this work, we aim to estimate the effect of individual variability on a set of very simple and easily interpretable features extracted from the EEG power spectra. In particular, in an identification scenario, we investigated how the aperiodic (1/f background) component of the EEG power spectra can accurately identify subjects from a large EEG dataset. The results of this study show that the aperiodic component of the EEG signal is characterized by strong subject-specific properties, that this feature is consistent across different experimental conditions (eyes-open and eyes-closed) and outperforms the canonically-defined frequency bands. These findings suggest that the simple features (slope and offset) extracted from the aperiodic component of the EEG signal are sensitive to individual traits and may help to characterize and make inferences at single subject level.

# Introduction

The scalp electroencephalogram (EEG) represents the most common recording system to detect human brain activity. Despite this technique dates back to more than a century ago, it still remains a clear reference in all the neuroscience related fields (Lopes da Silva 2013). One of the most relevant advantages of EEG, compared to other well-known techniques as Magnetoencephalography (MEG) and functional magnetic resonance (fMRI), is represented by its relative low-cost and its non-invasive properties that makes it especially suitable for real life applications. Merit of these characteristics, more recently, the use of scalp EEG has been extensively investigated as source of neurophysiological features to be used in the so called cognitive biometric systems (Chan et al. 2018; Gui et al. 2019). To date, an impressive number of feature extraction techniques have been used with the aim to detect subject-specific traits from EEG recordings. Time, frequency, time-frequency -domain techniques, functional connectivity metrics and the related derived network parameters have been extensively investigated into this specific research topic. Among these techniques, the most frequently used derive either from the use of power spectral density analysis at channels level (DelPozo-Banos et al. 2015; Rocca et al. 2014) or from functional connectivity between EEG channels (Barra et al. 2017; Crobe et al. 2016; Fraschini et al. 2015). All these approaches, from the most simple to the most sophisticated, are based on some sort of arbitrary choice, such as (i) the frequency band definition, (ii) the selection of a correlation metric for the connectivity and/or (iii) the application of a threshold to reconstruct the functional network. Interestingly, it has been shown that EEG brain activity exhibits a 1/f-like power spectrum (He 2014), defining an aperiodic component that may be characterized in terms of slope (i.e., the exponential decrease of power in a spectrogram as a function of frequency) and offset (i.e., offset of the broadband power of the signal). This arrhythmic brain activity has been associated with fluctuations of cognitive states (Podvalny et al. 2015), aging (Voytek et al. 2015), firing rate of neural populations (Manning et al. 2009) and clinical conditions (Peterson et al. 2018; Robertson et al. 2019; Veerakumar et al. 2019). Despite the interest on this specific approach, to the best of our knowledge, it has not been investigated yet to what extent this aperiodic component depicts the individual variability in EEG brain activity. In this study, we quantified the aperiodic component of the power spectrum, estimating the spectral slope and the offset at channel level. Successively, we used these EEG characteristics as feature vector to identify subjects in a large EEG dataset. Finally, we compared the results obtained from the aperiodic component with those obtained from more classical power spectral features, namely the relative power extracted from the canonical EEG frequency bands (i.e, theta, alpha, beta and gamma bands). In order to investigate this question, we used resting-state traces from a large and publicly available EEG dataset consisting of several recordings from 109 different subjects using a 64 channel EEG equipment (Goldberger Ary L. et al. 2000; Schalk et al. 2004).

## Material and methods

**Dataset**

To test our hypothesis we use a large and publicly available (https://physionet.org/content/eegmmidb/1.0.0/) EEG dataset consisting of several recordings from 109 different subjects using high-density EEG equipment (Goldberger Ary L. et al. 2000; Schalk et al. 2004). The EEG traces are provided in EDF+ format, containing 64 EEG signals and sampled at 160 samples per second. In particular, since the possible effects on individual variability of different sessions and different tasks have already been investigated (Fraschini et al. 2019; Pani et al. 2019) and are out of the scope of the present work, we focused our analysis on the two (one minute long) resting-state recordings during eyes-closed and eyes-open conditions.

**Preprocessing**

The preprocessing procedure was organized in two main steps. The first step was to apply ADJUST (version 1.1.1, https://www.nitrc.org/projects/adjust/), an automatic algorithm for artifact identification and removal (Mognon et al. 2011) based on Independent Component Analysis (ICA), with the aim to reduce the effects due to blinks, eye movements and other generic discontinuities. The subjects that showed significant residual artifacts were excluded from the analysis, in particular, the reported results are based on 95 subjects for the eyes-open condition and 100 subjects for the eyes-closed condition. The second step was to segment each (one minute) resting-state EEG recordings into five non-overlapping epochs of 12 seconds (Fraschini et al. 2015; Fraschini et al. 2016).

**Features extraction**

After the preprocessing steps, two different types of features were extracted from the epoched signals: (i) features characterizing the aperiodic component, namely the slope and the offset and (ii) features characterizing the periodic component, namely the relative power of theta [4-8 Hz], alpha [8-13 Hz], beta [13-30 Hz] and gamma [30-45 Hz] frequency bands. The slope and the offset were calculated using the Fitting Oscillations & One Over f (FOOOF) toolbox (Haller et al. 2018). The relative power for each frequency band has been computed as the ratio between the absolute band-specific power and absolute total power (between 1 and 45 Hz) using the Power Spectral Density estimate via Welch's method. Both approaches allowed to obtain, for each subject and each epoch, a features vector of 64 entries (each one representing the corresponding slope, offset or relative power of a single EEG channel).

**Performance evaluation**

To assess the performance of the two approaches (periodic and aperiodic components of the power spectra) to capture subject specific characteristics, we have tested the extracted features using a well known paradigm generally used to evaluate biometric systems (Fraschini et al. 2015). In particular, genuine and impostor scores were computed based on the Euclidean distance between pairs of feature vectors. Later, a similarity score was computed as $1/(1+d)$, where $d$ represents the Euclidean distance. Finally, the performance were derived from the false acceptance rate (FAR, the error that occurs when the system accepts an impostor) and the false rejection rate (FRR, the error that occurs when the system rejects a genuine match) at different thresholds. The area under the receiver operating characteristic (AUC) curves were evaluated together with the equal error rate (EER, the point where FAR equals FRR) and reported to summarize the results. Low values of EER and AUC express high performance. All the analysis was performed using Matlab (The MathWorks, Inc., Natick, Massachusetts, United States, version 2017B) and EEGLAB (version 13) (Delorme and Makeig 2004). All the figures were realized using Jamovi (version 1.0.8.0) available from https://www.jamovi.org.

# Results

**Eyes-open resting-state**

For the eyes-open resting-state condition, the best performance, in terms of EER and AUC, were obtained for the offset (EER = 0.079 and AUC = 0.025). The slope performed slightly worse, with EER = 0.090 and AUC = 0.031. Figure 1 represents the similarity score distribution for the slope (left panel) and the offset (right panel). When the two feature vectors were concatenated, the performance was slightly better with EER = 0.063 and AUC = 0.019. The relative power feature vectors did not show comparable performance, with the best results obtained for beta (EER = 0.118 and AUC = 0.043) and gamma band (EER = 0.112 and AUC = 0.039). All the results for eyes-open resting-state condition are summarized in Table 1.

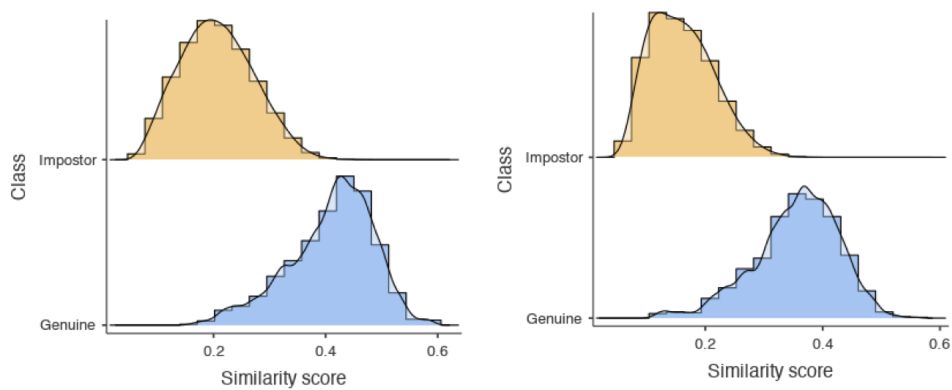

Figure 1. Similarity score distribution for the slope (left panel) and the offset (right panel)

| Feature | EER |
| --- | --- |
| slope | 0.090 |
| offset | 0.079 |
| slope + offset | 0.063 |
| theta relative power | 0.182 |
| alpha relative power | 0.256 |
| beta relative power | 0.118 |
| gamma relative power | 0.112 |

Table 1. A summary of the EER values for each feature vector.

**Eyes-closed resting-state**

For the eyes-closed resting-state condition, the best performance, in terms of EER and AUC, were obtained for the offset (EER = 0.057 and AUC = 0.018). The slope performed slightly worse, with EER = 0.089 and AUC = 0.035. Figure 2 represents the similarity score distribution for the slope (left panel) and the offset (right panel). In this case, when the two feature vectors were concatenated, the performance was slightly worse with EER = 0.171 and AUC = 0.019. Again, the relative power feature vectors did not show comparable performance, with the best results obtained for beta (EER = 0.142 and AUC = 0.078) and gamma band (EER = 0.197 and AUC = 0.124). All the results for eyes-closed resting-state condition are summarized in Table 2.

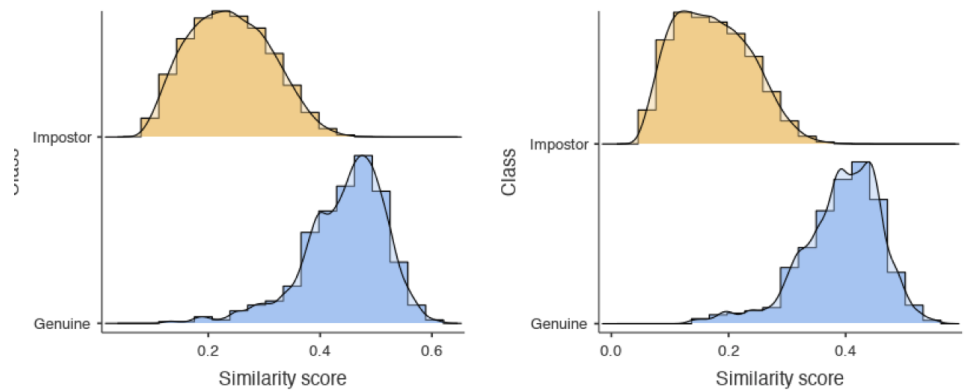

Figure 2. Similarity score distribution for the slope (left panel) and the offset (right panel)

| Feature | EER |
| --- | --- |
| slope | 0.089 |
| offset | 0.057 |
| slope + offset | 0.171 |
| theta relative power | 0.238 |
| alpha relative power | 0.181 |
| beta relative power | 0.142 |
| gamma relative power | 0.197 |

Table 2. A summary of the EER values for each feature vector.

**Discussion**

In this study, a large and open EEG dataset is used to investigate inter-subject variability using a fast and easy approach that goes beyond the traditional frequency band analysis. Two simple characteristics (slope and offset of the broadband power spectrum) describing the aperiodic properties of the EEG signals were exploited as fingerprints to identify individuals. Our results show that the aperiodic component of the power spectrum (i) is characterized by strong subject-specific properties, (ii) is consistent across different experimental conditions (eyes-open and eyes-closed) and (iii) its identification performance outperforms those obtained using canonical spectral features as band specific relative power.

Overall, the best performance is obtained using the aperiodic offset in the eyes-closed resting-state condition. Nevertheless, there is not a marked difference in terms of performance between experimental conditions (eyes-open versus eyes-closed resting-state) for any of the feature vectors tested. In line with previous studies (Crobe et al. 2016; M. Fraschini et al. 2015), for the periodic component we observed an effect on the performance due to the high frequency content, where for both conditions the lowest EER values (best performance) were obtained for beta and gamma frequency bands. The fact that the aperiodic component of the power spectrum outperforms the traditional spectral features (i.e., band specific relative power) is in line with the hypothesis that the a priori and arbitrary frequency bands definition and the averaging across these bands can mitigate individual variability (Haller et al. 2018). Furthermore, these results also support the hypothesis that this approach may help to better understand the role of oscillatory variability in explaining individual differences in cognitive functioning in health and disease (Haller et al. 2018).

Moreover, the aperiodic component has been linked experimentally to neuronal processes at micro-scale level (Manning et al. 2009; Miller et al. 2012; Podvalny et al. 2015) and it is hypothesized that its physiological meaning might reflect the dynamic balance between excitation and inhibition of neural population (Gao et al. 2017). In this work we tried to make a bridge testing if properties reported at micro/meso -scale were reliable at macroscale (i.e. EEG recordings). Similarly to our results, recent works at macroscale level showed the reliability of individual functional brain connectivity profile (Finn et al. 2015) and its strong genetic dependence (Demuru et al. 2017). A clear limitation of the present study is that it is based on a single session scenario and it is not clear how these findings may be altered using different EEG recordings in a multi-session scenario. Nevertheless, it is relevant to notice that our work is especially focused on relative (and not absolute) performance, since we present the results as a strict comparison between different characteristics of the EEG spectrum (namely, aperiodic and periodic components). Furthermore, the effects induced by different sessions on subject identification in EEG have been widely investigated and clarified (Gui et al. 2019) and out of the scope of the present study.

**Conclusions**

In conclusion, in this study we have shown that an EEG individual's profile, as defined by the aperiodic component of the power spectrum, is unique and it is possible to identify individuals (with very high accuracy) from a large EEG dataset. These findings suggest that these simple spectral features are sensitive to individual traits and may help to characterize and make inferences at single subject level using EEG.